\documentstyle[11pt]{article}
\date{}
\begin{document}
\sloppy
\title{The Energy Density of the Quaternionic Field as Dark
Energy in the Universe}
\author{V. Majern\'{\i}k \\
Institute of Mathematics, Slovak
Academy of Sciences, \\ Bratislava, \v Stef\'anikova  47,
 Slovak Republic\\ and\\
Department of Theoretical Physics\\ Palack\'y
University\\T\v r. 17. listopadu 50\\772 07 Olomouc, Czech Republic}
\maketitle
\begin{abstract}
\noindent
In this article we
describe a model of the universe consisting
of a mixture of the ordinary matter and a so-called cosmic quaternionic
field. The basic idea here consists in an attempt to interpret
$\Lambda$ as the energy density of the quaternionic field whose source is any
form of energy including the proper energy density of this field.
We set the
energy density of this field to $\Lambda$ and show that the ratio of
ordinary dark matter energy density assigned to $\Lambda$ is constant
during the cosmic evolution. We investigate the interaction
of the quaternionic field with the ordinary dark matter and show that
this field exerts a force on the moving dark matter which might
possible create the dark matter in the early universe.
Such
determined $\Lambda$ fulfils the requirements asked from the dark
energy.
In this model of the universe, the cosmical constant, the
fine-tuning and the age problems might be solved.
Finally, we sketch the evolution of the universe with the cosmic
quaternionic field
and show that the energy density of the cosmic quaternionic field
might be a possible candidate for the dark energy.
\end{abstract}
\section{Introduction}
According to the astronomical observation, the evidence continues to mount
that the expansion of the
universe is accelerating  rather than slowing down.
New observation suggests a
universe that is leigh-weight, is accelerating, and is flat
\cite{PER} \cite{pe} \cite{BO}.
To induce cosmic acceleration
it is necessary to consider some
components, whose equations of state are different from baryons,
neutrinos, dark matter, or radiation considered in the standard
cosmology.

As is well-known, in cosmology a new kind of energy is considered
called {\it quintessence} ("dark energy"). Quintessence represents
a dynamical
form of energy with negative pressure \cite{ST}.
The quintessence is supposed to obey an equation of state of the form
\begin{equation} \label{1}
p_Qc^{-2}=w_Q\varrho_Q,\qquad
-1<w_Q<0.
\end{equation}
For the vacuum energy
(static cosmological
constant), it holds $w_Q=-1$ and $\dot w_Q=0$.

An adequate cosmological theory, conform with
recent observation, should give answers to
the following problems \cite{S} \cite{P}:\\
(i){\it The cosmological constant problem.}
The '$\Lambda$-problem' can be expressed as discrepancies
between the negligible value of $\Lambda$ for the present universe and
the value $10^{50}$ times larger expected by Glashow-Salam-Weinberg model
\cite{WM}
or by GUT \cite{GUT} where it should be $10^{107}$ times larger.\\
{\it The fine-tuning problem}.
It is a puzzle why the densities of dark matter and dark energy are
nearly equal today when they scale so differently during the expansion
of the universe.
Assuming that the vacuum energy density is constant over time and the matter
density decreases as the universe expands it appears that their ratio
must be set to immense small value ($\approx 10^{-120}$) in the early
universe in order for the two densities to nearly {\it coincide} today, some
billions years later.\\
(iii) {\it The age problem}.
This problem expresses the discrepancy connected,
 on the one side,
with the hight estimates of the Hubble parameter and with the age of globular
clusters on the other side. The fact that the age of the universe is
smaller than the age of globular clusters is unacceptable.\\
(iv) {\it The flatness problem}.
Inflation predicts a spatially flat universe. According to Einstein's
theory, the mean energy density determines the spatial curvature of the
universe. For a flat universe, it must be equal to the critical energy.
The observed energy density is about one-third of critical
density. The discrepancy between the value of the observed energy
density and the critical energy is called the flatness problem.\\
(v) {\it Problem of the particle creation}.
In variable $\lambda$ models the creation of particles generally
takes place. The question what is the mechanism for this process
represents the problem of the particle creation.

It is well-known that the Einstein field equations with
a non-zero $\lambda$ can be rearranged so
that their right-hand sides consist of two terms:
the stress-energy tensor of the ordinary
matter and an additional tensor
\begin{equation}  \label{2}
T^{(\nu)}_{ij}=\left (\frac{c^4\lambda}{8\pi G}\right)
g_{ij}=\Lambda g_{ij}.
\end{equation}
$\Lambda$ is identified with vacuum energy
because this quantity satisfies the requirements asked from $\Lambda$, i.e.
(i) it should have the dimension of energy density, and (ii) it
should be invariant under Lorentz transformation. The second property
is not satisfied for arbitrary systems, e.g. material systems and
radiation. Gliner \cite{G} has shown that the
energy density of vacuum represents a scalar function of the
four-dimensional space-time coordinates so that it satisfies both above
requirements. This is why  $\Lambda$ is {\it identified}
with the vacuum energy.\\
From what has been said so far it follows that
the following properties are required from the vacuum energy density:
(i) It should be intrinsically relativistic quantity having the dimension
of the energy density.
(ii) It should be smoothly distributed throughout the universe.
(iii) It should cause the speedup of the universe.
(iv) It should balances the total mean energy density to $\Omega=1$.

In the next Sections we will
describe a model of the universe consisting
of a mixture of the ordinary matter and a so-called cosmic quaternionic
field.
The basic idea in this article consists in an attempt to interpret
$\Lambda$ as the energy density of the quaternionic field whose source is any
form of energy including the proper energy density of this field
The article is organized as follows. In Section 2 we describe
the proposed quaternionic field. In Section 3 we set the
energy density of this field to $\Lambda$ and show that the ratio of
ordinary dark matter energy density assigned to $\Lambda$ is constant
during the cosmic evolution. In Section 4 we show that this field exerts
force on moving dark matter. In Sections 5
we describe the possible
mechanism of the particle creation.
We sketch the evolution of the universe with the cosmic
quaternionic field.
Finally, we show that such
defined $\Lambda$ fulfils the requirements asked from the dark energy
and that in this model of the universe the above problems might be solved.

\section{The cosmic quaternionic field}

In a very recent article \cite{MK},
$\Lambda$ has been interpreted as the {\it field energy} of a
quaternionic field (called $\Phi$-field, for short)
\cite{MM} \cite{SIG} \cite{A} (see also the Appendix).
The field  equation
of the $\Phi$-field can be written
in the following form
\begin{equation}  \label{3}
\partial_i F_{ij}=J_j,
\end{equation}
where $J_i=k\rho v_i$ ( $v_i$ being the
4-velocity, $\rho$ matter density) is the current of the ordinary matter
and
$J_0=k(\epsilon_{self}+\rho)v_0$. $\epsilon_{self}$ is the energy
density of the $\Phi$-field and
$F_{ij}$ is the field tensor defined as
$$
F_{ij}=
\left( \begin{array}{cccc}
\Phi&0&0&0\\
0&\Phi&0&0\\
0&0&\Phi&0\\
0&0&0&-\Phi
\end{array} \right).
$$
$F_{ij}$ has only diagonal
components $F_{ii}=\Phi\quad i=1,2,3,\quad F_{ii}=-\Phi\quad
i=0,$ and $F_{ij}=0\quad
i\neq j$.  These components are transformed under Lorentz
transformation  as follows \cite{MA}
$$ F^{'}_{11}={1\over
{1-\beta^2}}(F_{11}+\beta^2F_{44})=F_{11}=\Phi,\quad
F^{'}_{22}=F_{22}=\Phi$$
$$F^{'}_{33}=F_{33}=\Phi \quad F^{'}_{44}={1\over
{1-\beta^2}}(F_{44}+\beta^2F_{11})=-\Phi=F_{44}.$$
The field
variable $\Phi$ has the dimension of field
strength and its square has the dimension of energy density.
Two scalars can be formed from $F_{ij}$: its trace
$F^{i}_{;i}$ and $F_{ij}F^{ij}$ the later has
the dimension of the energy density.
In the differential form the field
equations (3) are
\begin{equation} \label{5a}
\nabla \Phi=k \vec J= k\rho v_i,\quad\quad i=1,2,3
\end{equation}
and
\begin{equation}
-{1\over c}{\partial \Phi \over \partial
t}=k_0(\epsilon_{self}+\rho)v_i,
\quad\quad
i=0
\end{equation}
where $\epsilon_{self}$ is the energy density of the $\Phi$-field given
as \cite{MM}
$$\epsilon_{self}=\Phi^2.$$
From the Newton gravitation
law it follows that the gravitational "charge" of a point mass $m_{0}$
is $\sqrt
{G}m_{0}$ \cite{JM} \cite{B}.
Accordingly, we set for the
coupling constants $k$ and $k_0$
$$k=\frac{\sqrt{G}}{c}
\quad{\rm and}\quad k_0=\sqrt{G}.$$

These equations are first-order differential equations whose
solution can be found given the source terms.
Assuming the spacial homogeneity of the $\Phi$-field and the absence of
any
ordinary matter, i.e. $J_1= J_2= J_3=J_0=0.$ , the field variable $\Phi$
becomes independent of spatial
coordinates.
Here, the only source of the $\Phi$-field is its {\it
own} energy density, i.e. $\epsilon_{self}=\Phi^2$.
Therefore, Eqs.(4) and (5) become
\begin{equation}  \label{d}
\nabla\Phi= 0
\end{equation} \label{e}
\begin{equation} \label{f}
-{1\over c}{d \Phi \over d t}= {\sqrt {G}\Phi^2\over
c^2} ={\sqrt {G}\Phi^2 \over c^2}.
\end{equation}
The solution differential equation (7) has a
simple form
\begin{equation} \label{g}
\Phi(t)= {c \over \sqrt {G} (t+t_0)},
\end{equation}
where $t_0$ is the integration constant given by the boundary condition.
The energy density $\epsilon_{self}$ is approximately equal
to the observed value of the contemporary cosmological constant.

\section{The dark energy modeled by a time-dependent cosmological constant}
The theory of the time-dependent cosmological constant
in the Friedmann model is well
established ( see, e.g., \cite{AD}).
The time-dependent cosmological models in the framework of scalar field
theory were first discussed by P. J. E. Peeble and B.Ratra \cite{PR},
B.Ratra and P. J. E. Peebles (\cite{RA}) and M. \"Ozer
and M. O. Taha \cite{O}.
A number of authors set phenomenologically
$\Lambda\propto 1/t^2$ [22-29]
(for a review see \cite{OO}).
Generally, $\Lambda$  contains in its definition the gravitation
constant $G$ and velocity of light $c$.
From the purely phenomenological point of view
the simplest expression for $\Lambda\propto 1/t^2$, having the right
dimension,
and containing $G$ and $c$ is
$$\Lambda=\frac{\kappa^2c^2}{8\pi Gt^2},$$
where $\kappa$ is a dimensionless constant.
Because
$\Lambda$ is set equal to the {\it field energy density} of the cosmic
quaternionic field we have ($t_0=c=1$)
\begin{equation} \label{14}
\Lambda=\frac{1}{8\pi }\left [\frac{1}{\sqrt{G}t}
\frac{1}{\sqrt{G}t}\right ]
=\frac{\Phi^{2}}{8\pi}\qquad \Phi=\frac{1}{\sqrt{G}t},
\end{equation}
Accordingly, we have
\begin{equation}   \label{15}
\Lambda =\frac{\Phi^2}{8\pi}=\frac{1}{8\pi G t^2}\qquad {\rm
and}\qquad \lambda=\frac{1}{t^2}.
\end{equation}
The gravitational field equations with a
cosmological constant $\lambda$ and the
energy conservation law are (k=0)
\begin{equation} \label{16}
H^2=\frac{8\pi G}{3}(\rho+\Lambda)\qquad H=\frac{\dot R}{R}\qquad
\Lambda=\frac{\lambda}{8\pi G}
\end{equation}
\begin{equation} \label{17}
\frac{\ddot R}{R}=-\frac{4\pi G}{3}(\rho+3p+2\Lambda)
\end{equation}
and
\begin{equation} \label{18}
\dot \rho+3\frac{\dot R}{R}(p+\rho)=-\dot \Lambda.
\end{equation} \label{19}
Suppose we have a perfect-gas equation of state
\begin{equation} \label{20}
p=\alpha \rho
\end{equation}
and suppose that the deceleration parameter is constant.
If the evolution of the scale factor is given in form $R\propto t^n$ then
$q=-(n-1)/n$, therefore, we set
\begin{equation} \label{21}
q=-\frac{\ddot RR}{\dot R^2}=\frac{1}{n}-1.
\end{equation}
If we suppose the time dependence $\rho$ and $\Lambda$ in form
\begin{equation} \label{22}
\rho=\frac{A}{t^2}\qquad {\rm and} \qquad \Lambda=\frac{B}{t^2}\qquad
B=const. \qquad {\rm and}\qquad A=const.,
\end{equation}
respectively, then, inserting (16) into (12),(13) and (14), gives the
following relation between $A$ and
$B$ \cite{AB}
\begin{equation} \label{23}
2B=A[(-2+3n)(1+\alpha)].
\end{equation}
Given $A$ or $B$ and $n$ we can uniquely determine $B$ or $A$,
respectively.
For $\lambda \propto 1/t^2$, there is a relation between $\Omega_M$ and
the time-dependence of
scaling factor $R(t)$. Assuming that $\Omega_M$ does not change
during the matter-dominated era ($\alpha=0$) \cite{KM}
\begin{equation} \label{24}
R(t)=\left (\frac {3}{2}\right)^{\frac{2}{3\Omega_M}}(\Omega_M
C_1t)^{\frac{2}{3\Omega_M}}.
\end{equation}
The quantities $q$, $R(t)$ and $\Omega_M$ are mutually related.
Given one of them the remained quantities can be determined by means of
Eqs. (18), (16) and (15). It seems that $\Omega_M$ is best determined
by the observation,
therefore, we take it for the calculation of $q$ and $R(t)$.
Inserting $\Omega_M=1/3$ into Eq.(18) we obtain
$R\propto t^2 $ which yields $q=-1/2=1/n-1$, $n=2$.
We see from Eq.(9) that $B=1$ which inserting into Eq.(17) gives
$A= (1/2)$. The mean energy density $\rho$ and the cosmological
constant is given as ($\alpha=0$)
\begin{equation} \label{25}
\rho = \frac{1}{16\pi t^2}\qquad {\rm and}\qquad \lambda=\frac{1}{8\pi
G t^2},
\end{equation}
respectively.
Their ratio
\begin{equation}  \label{26}
\frac{\rho}{\Lambda}=1/2.
\end{equation}
Supposing the flat space, we have
\begin{equation}  \label{27}
\Omega_M=\frac{1}{3}\qquad {\rm and}\qquad
\Omega_{\Lambda}=\frac{2}{3}.
\end{equation}
There is no "fine tuning" problem in our model
since the ratio of the $\lambda$-part energy density to the mass-energy
density of the
ordinary matter remains during the cosmic evolution constant.

\section{The force exerting on the moving bodies in the cosmic
quaternionic field}

When the ordinary matter is present in the $\Phi$-field then
we have to apply the general field equations (3) and (5)
The ordinary matter changes
the value of the field variable $\Phi$ in comparison to the pure
$\Phi$-field.
Simultaneously, $\Phi$-field exerts a force $f_1$ on the moving matter.
This Lorentz-like force is given, similarly as in electrodynamics,
by the expression
\begin{equation}   \label{x}
f_j=\frac{J_i}{c}F_{ij}.
\end{equation}
When calculating this Lorentz-like force we take, for the save of
simplicity,
only the pure $\Phi$-field, i.e. that $\Phi$-field in which no
ordinary matter is
present.
$J_i^{(m)}=\frac{\sqrt
{G}}{c} m_0v_i$. Inserting $J_{i}^{(m)}$ in Eq. (22) we find for the
Lorentz-like  force
acting on the moving mass body in the presence of the $\Phi$-field
the following formula
\cite{X}.
\begin{equation} \label{28}
F_i=c^{-1}\sqrt{G}m_0\Phi v_i=  c^{-1}\sqrt{G}\Phi p_i.
\end{equation}
This is the fundamental formula which we apply in our further
consideration.

It is to be expected that the cosmic
$\Phi$-field manifests itself in the present-day
only in the following astrophysical situations:
(i) at the large mass concentrations, (ii) at the
large velocities of massive
objects
(iii) during the large time and space scales.
For the sake of simplicity, we confine ourselves to the non-relativistic
case, i.e.
we suppose that $m=$ const. and $v\ll c$. Then Eq.(6) turns out to be
\begin{equation}  \label{32}
 m\dot v =c^{-1}\sqrt{G}\Phi mv.
\end{equation}
Below, we present three possible effects of the quaternionic field in the
cosmic conditions:\\
(i) {\it The increase of the velocity of the moving bodies in
the $\Phi$-field}. Since $c^{-1}\sqrt{G}\Phi=1/t$ we get a
simple differential
equation $\dot v=\beta v $ where $\beta=1/t$
the solution of which is $v= Ct$.
A free moving object in the quaternionic field
is accelerated by a constant acceleration $C$. This
acceleration is due to the immense smallness of $\beta\approx 1/10^{18}$
in the present-day extremely
small. As is well-known for a given time instant the Hubble constant
$H$ is equal in
the whole Universe. Supposing $\beta|_{0}=H$,
the solution of equation $\dot
v= Hv$ becomes $v=Hr +C,$ where $C$ is an integration constant.
Setting $C=0$ we get a Hubble-like law $v=H r$.\\
(ii) {\it The increase of the kinetic energy of the moving bodies in the
$\Phi$-field.}
The gain of kinetic
energy of a moving body per time unit in the quaternionic field
is $(f_i\parallel v_i)$
\begin{equation} \label{33}
{dE\over dt}=F_iv_i = c^{-1}
\sqrt {G} \beta mv^2 =2 \sqrt{G}c^{-1}\Phi E_{kin}=2\beta E_{kin}.
\end{equation}
Again, the increase of the kinetic energy of a moving object is extremely
small. However, for a rapid rotating dense body it may represent a
considerable value.
For example, a pulsar rotating around its axis with the angular
velocity $\omega$ having the moment of
insertion $I$. Its kinetic energy is $E_{kin}\approx I\omega^2$ and its
change
in the quaternionic field is $dE_{kin}/dt\approx \beta I\omega^2\approx
10^{32}$ which is a value only of some orders of magnitude smaller than
the energy output of a pulsar \cite{HA}.\\
(iii) {\it The change of the kinetic parameters of the gravitationally
bounded moving bodies.} This
can be best demonstrated by describing
the motion of the Earth around the Sun taking into
account Eq.(13). It holds
\begin{equation}  \label{34}
\vec F_1+\vec F_2=-\frac{GM_{\odot}m_{\oplus}}{r^2} \vec r,
\end{equation}
where $\vec F_1=m_{\oplus}\vec {\ddot r}$ and $\vec F_2=
m_{\oplus}\vec {\beta \dot r}$.
Inserting $\vec F_1$ and $\vec F_2$ in
Eq.(26) we have
$$\frac{d}{dt}(\beta \dot {\vec r})=-\beta \frac{GM_{\odot}}{r^2} \vec r$$
from which it follows
\begin{equation}  \label{35}
\beta r^2 \dot \phi=const.=h
\end{equation}
The reciprocal radius $u$ satisfies the equation
$$\frac{d^2 u}{d \phi} +u=\beta^2 \frac{GM_{\odot}}{h^2},$$
the general solution of which is
$$u(\phi)=\frac{C}{t^2}+c_2\cos(\phi)-c_1 \sin(\phi)$$
Setting $c_1=c_2=0$, i.e. supposing that the orbit is circle we get
$$r\sim K t^2$$
Hence, the distance of the Earth and the Sun varies with time like
\begin{equation}  \label{36}
r\sim \frac{1}{GM_{\odot}\beta^2}\sim \frac{1}{\beta^2}\sim t^2.
\end{equation}
According to Eq.(28) the distance between the Earth and the Sun is
increasing direct proportional to the square of time.
There have been several atomic time measurements of the period of the Moon
orbiting around the Earth. A description of the work of some
independent research groups can be found in Van Flander's article
\cite{FL}. We simply point that after subtracting the gravitational
perturbative (tidal) effects, Van Flanders gives
$$\frac{\dot P}{P}=\frac{\dot n}{n}=(3.2\pm 1). 10^{-10}/{\rm yr},$$
where $n=2\pi /P$ is the angular velocity. Using Eq.(11) we get
comparable value
$$|\dot n/n|\approx 4.10^{10}/{\rm yr}.$$
However, given the complexity of the
data analysis, we must certainly await further confirmation by
different, independent test before concluding whether the $\Phi$-field
really affects the motion of cosmic bodies. Nevertheless, it can be
asserted that at present, there exists no evidence against the influence
of the $\Phi$-field on the moving bodies at the level of the supposed
present-day intensity of the $\Phi$-field.
Due to large value of the cosmic time the present-day effects of
the $\Phi$-field lei on the limit of the observability. However, they
had, probably, strong influence in the early universe . For
example, the strong $\Phi$-field can destabilize the large rotating mass
concentration (e.g. quasars) forming from them the present-day galaxies.
We note that the enlargement of
distance of the Earth and the Sun is also suggested by the large numbers
hypothesis presented by Dirac in 1937. In
this hypotheses Dirac supposed that $G\propto 1/t$. The astrophysical and
geological consequences of this hypothesis are discussed in details in
\cite{JO}\cite{WES}.

It is noteworthy that the force exerting on cosmical body in cosmic
quaternionic field is {\it always}
parallel to the direction of the velocity. This means that velocity of
moving bodies in cosmic $\Phi$-field is in all direction
increasing. For example, the moving bodies stemming from a exploding cosmic
body is equally speedup as those falling in the collapsing center.
 
\section{The creation of particle in the $\Phi$-field}

Another interesting effect of the cosmic $\Phi$-field is the
possibility of the creation of real particles from the virtual ones.
Particle creation in nonstationary strong fields is well-known phenomenon
studied
intensively in seventies ( see, e.g. \cite{GR}).
There are several proposed ways for the creation of real particles from
the virtual ones in the very strong and nonstationary gravitation field.
We propose here a new mechanism of the creation of real
particles from the virtual ones in the presence of the $\Phi$-field.\\
We note that our further
consideration on the creation of
matter from the vacuum quantum excitation are done by a semiclassic
way, although we realize that they should be
performed in terms of an adequate theory of
quantum gravity. However, as is well-known, when constructing a quantum
theory of gravity one meets conceptual and technical problems. The usual
concepts of field quantization cannot be simple applied to gravity
because standard field quantization (e.g., elm field) is normally done
in flat spacetime. It is impossible to separate the field equations and
the background curved space because the field equations determine the
curvature of spacetime. Moreover, the classical quantized field
equations are linear and that of gravitation are non-linear and
weak-field linearized gravitation field is not renormalizable.
There is even no exact criterion on which time and
space scales one has necessary to apply quantum laws for
gravitation field. Therefore, we take
the inequality $|A| \leq h $, where $A$
is the classical action, as a criterion for a possible application of
quantum physics in gravity.

According to
quantum theory, the vacuum contains many virtual particle-anti-particle
pairs whose lifetime
$\Delta t$ is bounded by the uncertainty relation $\Delta E\Delta t>h$
The proposed mechanism for the particle creation in the $\Phi$-field is
based on the force relation (23).
During the lifetime of the virtual particles
the Lorentz-like force (23) acts on
them and so they gain energy. To estimate this
energy we use simple heuristic arguments.
As is well-known, any virtual particle can only exist within limited
lifetime and its kinetics is bounded to the uncertainty relation
$\Delta p \Delta x>h$.
Therefore, the momentum of a virtual particle $p$ is approximately
given as $p\approx h\Delta x^{-1}$.
If we insert this momentum into Eq.(23) and multiply it by
$\Delta x$, then the energy of virtual
particle $\Delta E$, gained from the ambient $\Phi$-field during
its lifetime, is
\begin{equation}   \label{Q}
F\Delta x= \Delta E =\sqrt{G} \Phi(t){h\over c}.
\end{equation}
When the $\Phi$-field is sufficiently strong then it can supply enough
energy to the virtual particles during their lifetime and so
spontaneously create real particles
from the virtual pairs. The energy necessary for a particle to be created
is equal to $m_vc^2$ ($m_v$ is the rest mass of the real particle).
At least, this
energy must be supplied from the ambient $\Phi$-field to a virtual particle
during its lifetime. Inserting $\Phi$ into Eq. (29), we have
$$\Delta E\approx {h\over (t+t_0)},$$
Two cases may occur:
(i) If $m_vc^2<\Delta
E$, then the energy
supplied from the $\Phi$-field is sufficient for creating real particles of
mass $m_v$
and, eventually, gives them an additional kinetic energy.
(ii) If $m_v>c^2\Delta
E$, then
the supplied energy is not sufficient for creating the real particles  of
mass $m_v$ but only the energy excitations in vacuum.\\
The additional kinetic
energy of the created particles, when $\Delta E>m_vc^2$, is
$$E_{kin}=\Delta E -m_vc^2={h\over (t+t_0)} - m_vc^2.$$
During the time interval
$(\approx 0,10^{-20})$ after the Big Bang, the masses of
the created particles lei in the
range from $10^{-5}$ to $10^{-27}$ g.
Their kinetic energy
was $E_{kin}=[h/(t+t_0)]-m_0c^2$.
reached values up to $10^{-5}$ erg, which corresponds to
the temperature of $10^{21}$ K.
Today, energies of the virtual pairs, gained during their lifetime,
are negligible small.

\section{The energy density of the cosmic $\Phi$-field as a possible
candidate for the dark energy}

As has been shown in the previous Chapters,
when taking the field energy density of cosmic quaternionic
field as the vacuum
energy density, the problems presented in the Introduction may be
resolved. The problem of the
cosmological constant, because the value of $\Lambda$ is consistent with data,
the tuning and age problems because the ratio of mass to vacuum energy
density does not
vary during the cosmical evolution and
the age of the universe is large
enough to evolve the globular cluster.
The flatness problem is also solved because the sum of the dark energy
to the ordinary matter yields just the critical energy density.
Moreover, in the cosmic quaternionic field,
there exists a plausible mechanism of
the particle creation.
We conclude that the energy density of the $\Phi$-field represent
a relativistic quantity satisfying Gliner's requirements which is smoothly
distributed in space. It causes the speedup of the universe,
balances the total energy density to the critical one and
gives a plausible mechanism for the particle creation.

It is generally believed that dark energy was less
important in the past and will become more important in the future.
In our model the value of dark energy is proportional to $1/t^2,$
therefore, its value becomes very large in past and will be adequate
large in the future. We remember that the force exerted by the cosmic
$\Phi$-field on moving bodies acts always in the direction of the
velocity. That means that the high value of the dark energy in the
early universe does not interfere with the structure forming, contrarily,
it accelerates it.

The evolution of the universe with the cosmic quaternionic field
can be briefly sketched as follows:
The cosmic evolution started purely
field-dominated era with the inflation, after which
a massive creation of particles began together with enormous release of
entropy. The masses of the created particles reaches values up to $-5$
g and the kinetic energy
of the created nucleons
values up to $10^{-5}$ erg, which corresponds to
the temperature of $10^{21}$ K.
The large vacuum energy density of the cosmic quaternionic field at the
early stage of the universe
accelerates its structure formulation. From what has been said above
we conclude that the energy density of the cosmic
quaternionic field
might be a possible candidate for the dark energy because (i) it has the
value consistent with data (ii) it does not suffer from the cosmological
constant, fine-tuning, age and
flatness problems (iii) it yields a plausible mechanism particle production
and (iv) it accelerated the structure formulation in the early universe.

Motivated by the desire to find a possible candidate of the dark energy
among the family of quaternionic fields we found the suprisingly simple
quaternionic field whose energy density
might be considered as the dark energy. (This points out that also the
classical field may be interested by study of the quintessence \cite{Va}.
The energy density of this
field has the desired properties of dark energy and changes generally
our view of the vacuum energy density modelled by the cosmological
constant. It could not be
seen as the carrier of repulsive gravity but as an amplifier of velocity
of the moving bodies independently of the direction of their motion.
The fact that the
vacuum density accelerates the expanding galaxies is cause do to fact that
they move from the center of the universe.
When the galaxies would
collapse the vacuum energy would accelerate, likewise, their
collapsing.
This may eventually lead to change the basic equation of the Friedmann
cosmology.
As well-know there are many versions of Mach's principle
in the literature and a
unique, satisfactory formulation of this principle does not seem
to exist as yet.
It seem that the influence of the moving bodies changes the
value the field variable $\Phi$ which for its part
affects other cosmic bodies might be seen as a form of Mach's principle.

\vspace{2cm}
\centerline{\bf APPENDIX}
\vspace{0.4cm}

The quaternionic field equations can be described by a quaternionic
equation consisting
of the quaternionic differential operator, the field
and source quaternions \cite{MM}.\\
(i) The quaternionic differential operator is the quaternion ($i,j,k$ are
the quaternionic units obeying the following relations $ij=-ji,\quad
ik=-ki,\quad jk=-kj$ and $i^2=j^2=k^2=-1$) \cite
{MM} \cite{A}
$$ \mbox{\fbox{$\cdot$}} =i{\partial \over \partial x} +j{\partial \over
\partial y} + k{\partial \over \partial z} +{s\over c}{\partial \over
\partial t}\quad s=\sqrt{-1}.\quad \eqno (A1)$$
(ii) The field quaternion the components of which are the field
variables is
$${\bf \Phi} = i\Psi_1+j\Psi_2+k\Psi_3 +\Psi_4,\quad \eqno(A2)$$
where
$$\Psi_1=(\Phi_1+s\Phi_1^{'}),\quad \Psi_2=(\Phi_2+s\Phi_2^{'}),
\quad\Psi_3=(\Phi_3+s\Phi_3^{'}),\quad \Psi_4=
(\Phi_4 +s\Phi_4^{'}).$$
(iii) The source quaternion
$$J={4\pi \over
c}[iJ_1+jJ_2+kJ_3+ J_4] ,\quad \eqno (A3)$$
where
$$J_1=(J_x^{'}-sJ_x),\quad J_2=(J_y^{'}-sJ_y)\quad J_3=(J_z^{'}-sJ_z)],\quad
J_4=j_0
+sj_0{'}.
 $$
The general  quaternionic field is described by the field  equation
of the type (see, e.g. \cite{SIG})
$$\stackrel{\longrightarrow }{\mbox{\fbox{$\cdot$}}}\ \Phi =
 J \quad \eqno (A4)$$
The field quaternion ${\bf \Phi}$
consists of the vector part $\vec \Psi_1=
\vec \Phi_1+s\vec
\Phi_2$, where $\vec
\Phi_1=(\Phi_1,\Phi_2,\Phi_3)$, $\vec \Phi_2=(\Phi_1,\Phi_2,\Phi_3)$ and
the scalar part
$\Psi_4=(\Phi_4+s\Phi_4)$.\\
The source quaternion consists likewise of the vector part
$\vec J =\vec J_1-s\vec J_2$, where $\vec
J_1=(J_x^{'},J_z^{'},J_z^{'})$ and $\vec
J_2=(J_x, J_y, J_z)$ and the scalar part $J_4=4\pi (j_0+ sj_0^{'})$.\\
The energy density of a quaternionic field is \cite{MM}
$${\bf E}= \sum_{k=1}^{4}\Phi_i\Phi_i^{*}.\quad\eqno (A5)$$
By using the vector notation we can rewrite the quaternionic field
equations (A4) in the form
$${s\partial \Psi_4\over c\partial t} +\nabla. \vec \Psi = 4\pi J_{4} $$
$${s\partial \vec \Psi \over c\partial t} -\nabla \Psi_4+\nabla\times
\Psi= {4\pi\over c}\vec J.\quad \eqno (A6) $$

According to the specification of the field variables and source components
in the field and source quaternion, respectively, we get the following
fields:\\
(i) If we choose $\Phi=- \vec E, \quad \Phi^{'}= - \vec B, \quad \vec J =-\vec
J\quad j_0=-\varrho $,
$\Phi_{4}=\Phi_{4}^{'}=
J^{'}=j_{0}^{'}=0 $ and we associate
$J,j_{0}$ with the components of electromagnetic 4-current,
we get just the standard Maxwell equations
\cite{E} \cite{Ed2}. \\
(ii) If we associate, in addition, $J^{'}$
and $j^{'}$ with the components
of the monopole
4-current then we
get the Maxwell equations with monopoles \cite{BL}.\\
(v) If we associate the components of the field quaternion with the field
variables as in (i) and take
$J,j_{0}$ as electric and $J^{'},j^{'}_{0}$ as monopole currents, $\Phi_{4}$
and
$\Phi_{4}^{'}$ as the scalar and the pseudoscalar variables,
respectively,
then
we obtain the Ohmura field  equations \cite{O}.\\
(vi) If we put
$\vec \Phi =\vec \Phi^{'}=\Phi{'}_4=\vec J'=j_0'=0$, and take $J^{*},j^{*}$
as not specified 4-current, we get the scalar quaternionic field equations.\\
We note that all of these field equations can be written also in form of
tensor equation by means of the corresponding field tensors \cite{MM}.
The scalar quaternionic field represents the $\Phi$-field, whose
field equations in vector form we consider in Chapter 2.\\

\end{document}